\documentclass[12pt]{iopart}
\usepackage{amssymb,epsfig,graphicx}

\begin{document}

\title{Transverse ratchet effect and superconducting vortices:\ Simulation
and experiment}
\author{L Dinis$^{1}$, D Perez de Lara$^{2}$, E M Gonzalez$^{2}$, J V Anguita%
$^{3}$, J M R Parrondo$^{1}$ and J L Vicent$^{2}$}

\begin{abstract}
A transverse ratchet effect has been measured in magnetic/superconducting
hybrid films fabricated by electron beam lithography and magnetron
sputtering techniques. The samples are Nb films grown on top of an array of
Ni nanotriangles. Injecting an ac current parallel to the triangle
reflection symmetry axis yields an output dc voltage perpendicular to the
current, due to a net motion of flux vortices in the superconductor. The
effect is reproduced by numerical simulations of vortices as Langevin
particles with realistic parameters. Simulations provide an intuitive
picture of the ratchet mechanism, revealing the fundamental role played by
the random intrinsic pinning of the superconductor.
\end{abstract}
\pacs{05.40.-a, 02.30.Yy,74.25.Qt, 85.25.-j}

\address{$^{{1}}$ Grupo Interdisciplinar de Sistemas Complejos (GISC)
and Departamento de F\'isica At\'omica, Nuclear y Molecular. Universidad
Complutense de Madrid. E-28040 Madrid, Spain.\\ $^{{2}}$ Departamento de
F\'isica de Materiales. Universidad Complutense de Madrid. E-28040 Madrid, Spain.\\
$^{{3}}$ Instituto de Microelectr\'onica de Madrid. Consejo Superior
de Investigaciones Cient\'ificas. Tres Cantos, E-28760,Spain.}

\maketitle

\section{Introduction}

The study of transport in asymmetric substrates, under the generic name of
``ratchet phenomena'', has attracted increasing attention in the last years.
Ratchets exhibit unexpected transport properties, such as rectification or
negative resistance, which can be used to control motion and also to reveal
aspects of the underlying dynamics in many physical systems \cite%
{reimann,Linke_App}. This is the case of vortex motion in superconductors,
where different types of ratchets have been implemented using substrates
with asymmetric defects \cite{Science}.

\begin{figure}[tbp]
\centering
\includegraphics[width=0.8\textwidth]{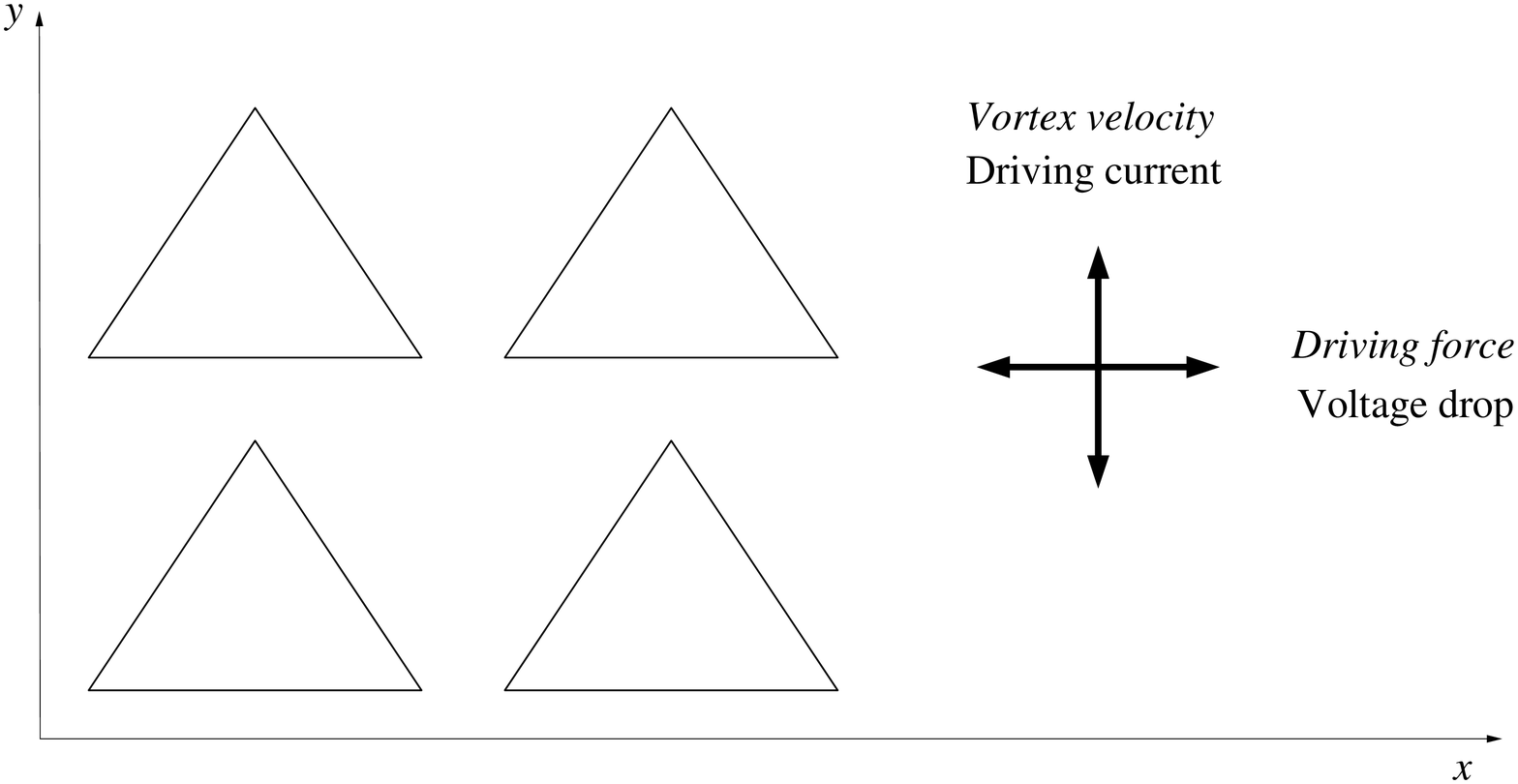}
\caption{Transverse ratchet configuration.}
\label{scheme}
\end{figure}

In ratchets, asymmetry is normally used to rectify an ac signal, but it can
also induce other non trivial transport effects. One of these effects is the
so-called transverse ratchet, in which an ac force can induce a directed
motion \emph{perpendicular} to the force. Several authors have theoretically
worked on this topic \cite{Derenyi,Bier,Savelev,kolton}. Olson-Reichhardt
and Reichhardt \cite{reichhardt_a} have studied this effect by numerical
simulations of superconducting vortices as Langevin interacting particles.
In a recent publication, Gonzalez \textit{et al.} \cite{vicent_trans} have
presented experimental evidence of transverse ratchet in superconducting
samples, using non-superconducting triangles embedded in a superconducting
film. In the usual ratchet effect configuration the driving current is
applied perpendicular to the triangle reflection symmetry axis (tip to base
axis) and the output dc voltage signal is measured on the same direction.
However, in the case of transverse ratchet, the driven current is applied
parallel to the triangle reflection symmetry axis, and the output voltage
drop is measured perpendicular to the input current direction, i.e.
perpendicular to the triangle reflection symmetry axis. We recall that the
voltage drop in one direction probes the vortex motion along the
perpendicular direction, as given by the Lorentz force and the Josephson
equation \cite{vicent_trans} (a schematic representation of the transverse
setup is shown in figure \ref{scheme}).

In this work we show that this transverse ratchet effect can be modeled in
the framework of the Langevin equation in two dimensions (2D), taking into
account that we are dealing with an adiabatic ratchet effect of interacting
particles \cite{vicent_PRBRatchet}. The crucial point is the interplay
between two pinning potentials: i) intrinsic and random pinning potentials,
due to the structural defects present in the superconducting films, ii)
artificial periodic ratchet pinning potentials, due to the array of non
superconducting nanostructures embedded in the superconducting films. This
approach is able to reproduce the sign and magnitude of the experimental
data, with realistic values of all the parameters involved in the simulation.

\section{Experimental method}

The samples are superconducting/magnetic hybrids; i.e. Nb films grown on top
of arrays of Ni nanotriangles which are fabricated on Si (100) substrates.
These samples are obtained following several steps with different
techniques. The first one is e-beam writing of the nanotriangles on a
polymethyl methacrylate (PMMA) resist covering the Si (100) substrate, next
developing using methyl isobutyl ketone : isopropyl alcohol (1:3) during 15
seconds, and magnetron sputtering deposition of Ni. Once lift-off is
performed and the resist is removed, only nanometric Ni triangles remain on
top of the substrate, which is then covered by a thin film of Nb, also by
magnetron sputtering technique. The thicknesses are always the same, for Ni
(triangles height) 40 nm, and 100 nm for the Nb film. The nanotriangles side
is around 600 nm. The array period is $\mbox{770 nm}\times \mbox{750 nm}$.
Samples were lithographed with a cross-shaped bridge (40 $\mu $m wide) 
using standard optical lithography and reactive ion etching techniques for
magnetotransport measurements. This cross-shaped bridge allows injecting a
transport current and measuring voltage drops along two perpendicular
directions. This guarantees that the only asymmetry in the experimental
layout is coming from the Ni triangular traps. Magnetotransport measurements
were carried out in a commercial liquid He cryostat provided with a
superconducting magnet and a variable temperature insert. The magnetic
field, which creates the vortex lattice, is applied perpendicular to the
film. The magnetic properties of these arrays of Ni triangles have
been already published \cite{jaafar}. The temperature is kept constant
and close to the superconducting critical temperature, which is around 8.7 K
in our samples. The measurements have been done with a frequency of the ac
applied current of 10 kHz. The experimental configuration is Hall-like, the
injected ac current is applied parallel to the triangle reflection symmetry
axis (tip to base axis) and the output voltage is recorded perpendicular to
this axis. This type of measurements could yield experimental artefacts due,
for instance, to misalignment of the potential contacts. In the related
literature one can find well known experimental methods to avoid these
unwanted effects for all kind of unpatterned \cite{colino} or patterned \cite%
{crusellas} samples. In our case, the experimental signal coming from these
effects is much smaller than the transverse ratchet effect signal and the
possible contribution to the experimental data could be neglected as was
reported in \cite{vicent_trans}.

Finally, close to critical temperature \cite{martin_velez}
magnetoresistance of superconducting thin films with periodic arrays of
pinning centers show minima when the vortex lattice matches the unit cell of
the array \cite{martin}. This geometric matching occurs when the vortex
density is an integer multiple of the pinning center density, hence the
number $n$ of vortices per array unit cell can be known by simple inspection
of the magnetoresistance curves, in which the first minimum corresponds to
one vortex per unit cell, the second minimum to two vortices per unit cell,
and so on. Therefore, changing the applied magnetic field we can select the
number of vortices per array unit cell.

The transverse ratchet effect has been observed for different values of $n$,
as shown in Fig.~\ref{fig_results}. 
Firstly, we proceed to describe the numerical simulations and then we
discuss the results.

\section{Numerical simulations}


We have performed extensive numerical simulations of vortices as a set of 2D
interacting, overdamped Brownian particles in the Langevin approach. This
type of simulations have been used to study rectification, current reversal
and lattice configurations effects \cite{reichhardt_a,dinisn3,dinisn4}.
Transverse vortex rectification has been previously analyzed by Olson-
Reichhardt, and Reichhardt \cite{reichhardt_a} and Savel'ev \emph{et al.} 
\cite{Savelev} using similar simulations. 
However, our experimental results differ from these simulations. 

Even though random intrinsic pinning effects in Nb and type-II
superconductors without periodic pinning traps have been widely discussed 
\cite{campbell}, Langevin type simulations of experimental vortex ratchet
systems usually disregard the effect of intrinsic pinning centers \cite%
{reichhardt_a,dinisn3,dinisn4,souza_silva,lu,zhu_physe,reichhardt_b}. As
pointed out by A.~Kolton~\cite{kolton}, pinning defects may have a strong
influence in the transversal ratchet effect. In our simulations Nb intrinsic
pinning is taken into account as a random distribution of potential wells,
and as we will see later, the disorder induced by the intrinsic pinning in
the vortex lattice plays a fundamental role in the transverse ratchet effect.

Langevin equation for the position $\mathbf{r}_i$ of the $i$-th vortex
reads: 
\begin{equation}
\eta \dot{\mathbf{r}_i}(t)=-\sum_{j\neq i}\vec{\mathbf{\nabla}}%
_{i}U_{vv}\left(\left|\mathbf{r}_i-\mathbf{r}_j\right|\right)-\vec{\mathbf{%
\nabla}}_i{V}_{tp}(\mathbf{r}_i)-\vec{\mathbf{\nabla}}_i{V}_{ip}(\mathbf{r}%
_i)+\mathbf{F}_{\mathrm{ext}}(t)+\mathbf{\Gamma}_i (t)  \label{eq:langevin}
\end{equation}
where $\eta$ is the viscosity, $U_{vv}(r)$ is the interaction potential
between vortices, $\mathbf{F}^{tp}_i(\mathbf{r}_i)=-\vec{\mathbf{\nabla}}_i{V%
}_{tp}(\mathbf{r}_i)$ the force due to Ni triangular pinning traps, $\mathbf{%
F}^{ip}_i(\mathbf{r_i})=-\vec{\mathbf{\nabla}}_i{V}_{ip}(\mathbf{r}_i)$ the
force due to randomly distributed pinning centers present in the Nb sample, $%
\mathbf{F}_{\mathrm{ext}}(t)$ the external force due to the applied current,
and $\mathbf{\Gamma}_i(t)$ are white Gaussian noises accounting for thermal
fluctuations: 
\begin{equation}
\langle \mathbf{\Gamma}_i(t)\cdot\mathbf{\Gamma}_j(t^{\prime })\rangle ={4kT}%
{\eta}\delta_{ij} \delta (t-t^{\prime }).
\end{equation}

For the vortex interaction potential, we have used \cite{tinkham}: 
\begin{equation}
U_{vv}(r)=\frac{\phi _{0}^{2}d}{2\pi \mu _{0}\lambda ^{2}}K_{0}\left( \frac{r%
}{\lambda }\right)  \label{eq:vv}
\end{equation}%
where $\phi _{0}=h/(2e)$ is the quantum of flux, $\mu_0$ is the magnetic
permeability of vacuum, $K_0$ is the zeroth-order modified Bessel function, $%
\lambda$ is the penetration depth of the material, and $d$ is sample
thickness.

For the friction coefficient $\eta$, we have used Bardeen-Stephen
contribution to vortex viscosity per unit length giving $\eta=\phi_0^2d/2\pi%
\xi^2\rho_n$, where $d$ is the sample thickness, $\xi$ the vortex coherence
length and $\rho_n$ is normal state resistivity for Nb.

The external force is sinusoidal as in the experiment, with the same
amplitude but different frequency. Friction coefficient and vortex
interaction strength set the appropriate time scale for simulation, and
unfortunately, the small time step required ($\Delta t=2\times10^{-6}\mu$s)
prevents us from using the experimental value of applied force frequency of $%
10\mbox{kHz}$. However, experiments show that the system is adiabatic and
the result does not depend on signal frequency, which allows us to use an
adiabatic approximation as in \cite{dinisn4}. Finally, we have compared
adiabatic simulations with those using an AC signal of $5\mbox{MHz}$ showing
that this is a sufficiently small frequency as to yield the same results
(not shown). For the rest of the simulations we have used a $5\mbox{MHz}$
signal because this slightly simplifies data analysis by avoiding numerical
computations of the adiabatic integral. 
%

\begin{figure}[tbp]
\centering
\includegraphics[width=0.8\textwidth]{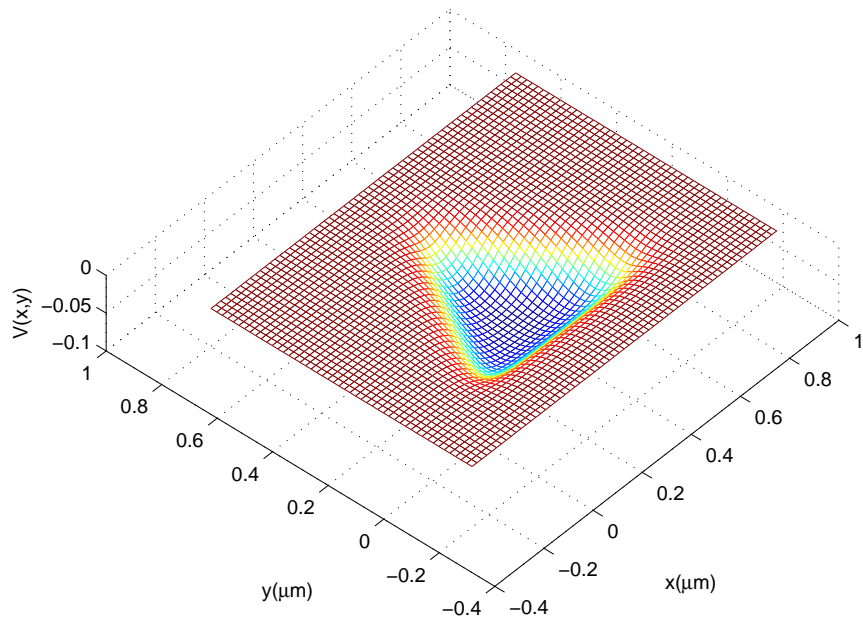}
\caption{Potential used for vortex-triangle interaction.}
\label{fig_potencial_triangulo}
\end{figure}

The interaction between vortices and the triangular Ni pinning wells is
modeled by a potential $V_{tp}$ in the shape of a triangle and with a smooth
hyperbolic tangent profile and rounded vertices and intensity $V_{tp0}$, as
depicted in figure \ref{fig_potencial_triangulo}.

Intrinsic pinning defects have been simulated as a series of potential wells
randomly distributed across the sample and in the shape of paraboloids
truncated at radius $r_d$. Thus, the interaction potential of vortex $i$
with a defect in position $\mathbf{r_c}$ is 
\begin{equation}
V_{ip}(\mathbf{r}_{i})=-V_\mathrm{def}+\frac{V_\mathrm{def}}{r_d^2}(\mathbf{r%
}_{i}-\mathbf{r}_{c})^2, \mbox{ if } \left|\mathbf{r}_i-\mathbf{r}%
_c\right|<r_d
\end{equation}
and zero otherwise, where $V_\mathrm{def}$ represents the intensity of the
intrinsic pinning potential. Following \cite%
{jensen,dong,wang_a,wang_b,zhu_physc}, we have chosen the size of the defect
as the size of the vortex core $r_d=\xi$, which corresponds to point defects
interacting with the core of the flux quantum. If the density of pinning
defects is sufficiently high as it is in Nb samples 
the overlapping pinning centers will lead to a diffuse potential or
``pin-scape'' which can be described by a much lower effective pinning
density and certain amplitude \cite{jensen}. Only a small fraction of the
resulting wells will have an amplitude large enough to produce appreciable
pinning, the majority of them overlap producing a low amplitude wriggling in
the pinning potential. This effective density has been chosen large enough
so that the defects produce lattice distortion, but not too large so that
computer simulation time remains below a reasonable limit. 

Table~\ref{tab_param} presents a summary of numerical values used in
simulation.

\begin{table}[tbp]
\centering
{\footnotesize \textrm{%
\begin{tabular*}{\textwidth}{@{}c@{\extracolsep{0pt plus12pt}}c@{\extracolsep{0pt plus12pt}}c@{\extracolsep{0pt plus12pt}}c@{\extracolsep{0pt plus12pt}}c@{\extracolsep{0pt plus12pt}}c@{\extracolsep{0pt plus12pt}}c@{\extracolsep{0pt plus12pt}}c@{\extracolsep{0pt plus12pt}}c@{\extracolsep{0pt plus12pt}}c@{\extracolsep{0pt plus12pt}}c@{\extracolsep{0pt plus12pt}}c@{\extracolsep{0pt plus12pt}}c@{\extracolsep{0pt plus12pt}}c@{\extracolsep{0pt plus12pt}}c}
\br Magnitude & Symbol & Value   \\
\mr Friction coefficient * \cite{bardeen_stephen} & $\eta$  & $6\times 10^{-5}$ pN$\mu $s/$\mu$m   \\
\mr Penetration length * \cite{hake} & $\lambda$ & 0.320 $\mu$m  \\
\mr Coherence length * & $\xi$ & $0.090$ $\mu$m  \\
\mr Sample thickness *& $d$ & $0.100$ $\mu$m  \\
\mr Temperature * & $T$ & $8.3$ K \\
\mr Triangular pinning **& $V_\mathrm{tp0}$ & $0.08$ pN$\mu$m  \\
\mr Intrinsic pinning **& $V_\mathrm{def}$ & $0.015$ pN$\mu$m   \\
\mr Defect density **& $\rho_\mathrm{eff}$ & $14.5$ $\mu$m$^{-2}$   \\
\br
\end{tabular*}
}}
\caption{Numerical values used in simulations in units pN, $\protect\mu$m, $ \protect\mu$s and K. *: Experimental values. **: Values that have been adjusted to reproduce experimental data behaviour.
\label{tab_param}}

\end{table}

\begin{figure}[tbp]
\includegraphics[width=0.5\textwidth]{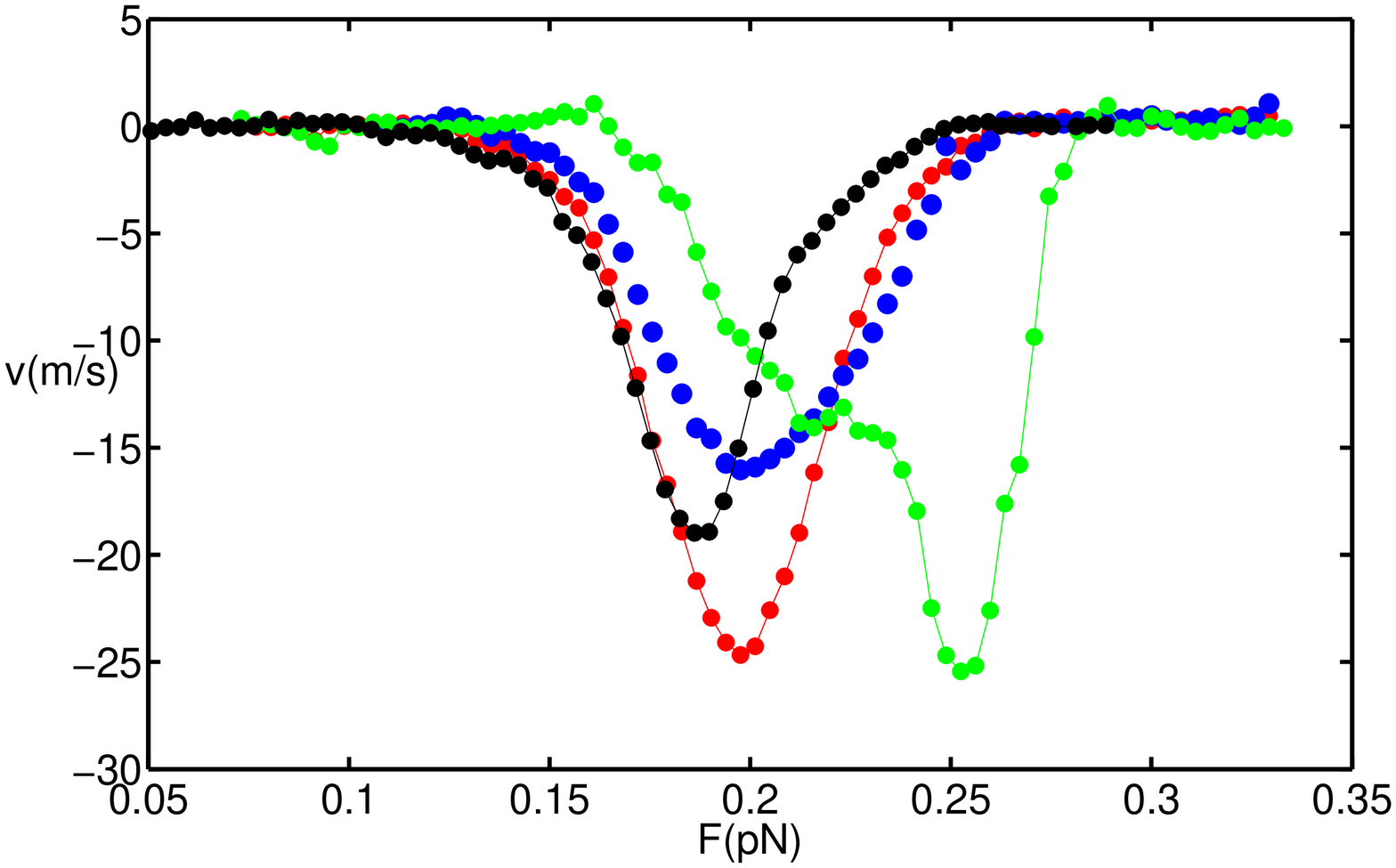}
\includegraphics[width=0.5\textwidth]{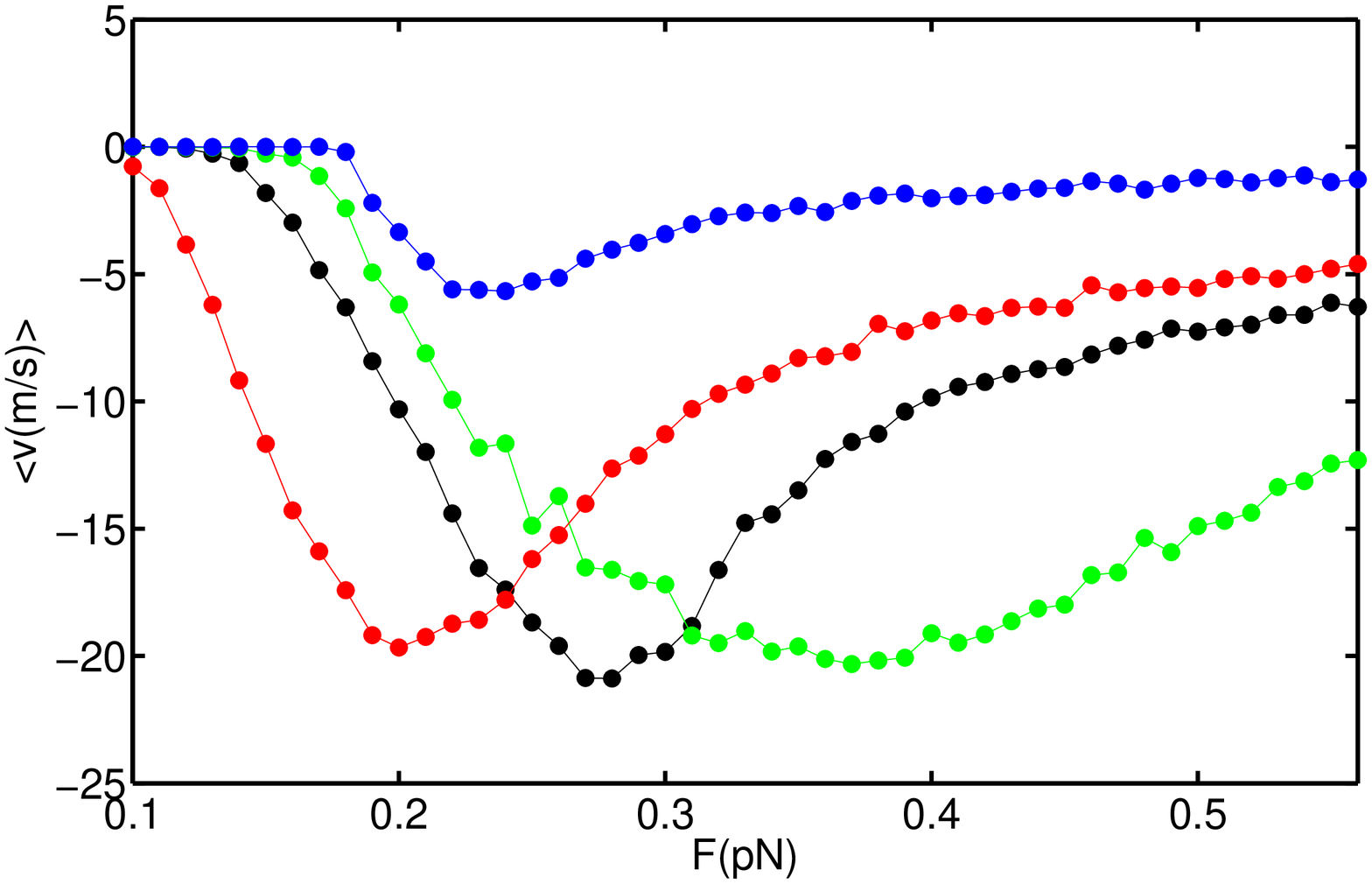}
\caption{\emph{Left:} Measured average vortex speed versus amplitude of
applied force, for applied magnetic field $H=64$Oe ($n=2$) (green), $H=96$Oe ($n=3$) (black), $H=128$ Oe
($n=4$) (blue) and $H=176$ Oe ( $n=6$ )(red), $T/T_c=0.98$. \emph{Right%
}: Simulation results, average vortex speed versus amplitude of applied
force, for $n=2$ (green), $n=3$ (black), $n=4$ (blue) and $n=6$ (red) vortices per unit cell. The lines joining the dots is only a guide for the eye. }
\label{fig_results}
\end{figure}

Experimental and simulation results are represented in figure~\ref{fig_results}.
Simulations and experiment data show similar behavior. Several
differences can be observed though. For instance, simulated vortex velocity does
not decay with force as fast as in the experimental curves. In the case of the simulations, the decay is that of an adiabatic system where the velocity is expected to decay slowly, as the
inverse of the square root of the applied peak force intensity. This
is due to the fact that even for very high peak forces, the sinusoidal
applied force covers small values of the force for a short interval.
This is however not observed in the experimental data although the
system is known to be adiabatic with respect to the applied force
frequency \cite{Science,vicent_PRBRatchet}.
At high enough driving forces, a vortex velocity threshold is
found in the experiments above which the interaction between the moving vortex lattice
and the ordered array is very weak, if any, and commensurability effects are
suppressed as was reported by Velez et al.~\cite{velez_jaque}. Nevertheless, we point out that the sign of the current is correctly reproduced by our simulations, which also agree with the order of magnitude of
velocities and applied force.

\section{Discussion: rectification mechanism}

\begin{figure}[tbp]
\centering
\includegraphics[width=0.4\textwidth]{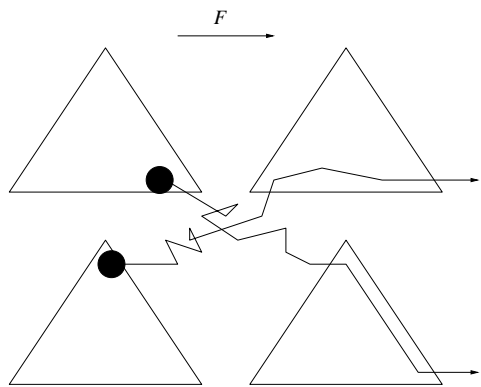}
\caption{Sketch of rectification mechanism (see text for discussion).}
\label{fig_mechanism}
\end{figure}

The mechanism for transverse rectification is depicted in figure \ref%
{fig_mechanism} (see also supplementary material) and can be stated as
follows:

\begin{enumerate}
\item Fluctuations may take a vortex to a neighbouring row of triangles.

\item If a vortex enters a triangle by its tip (downward motion in figure %
\ref{fig_mechanism}) it may be carried further down parallel to the side of
the triangle by the combination of the horizontal force and the triangular
potential. If the vortex enters the triangle at its base (upward motion in
figure \ref{fig_mechanism}), the external force and the triangle potential
keep it close to the triangle base. Thus, downward fluctuations are promoted
or favoured, ratcheting the particles down and yielding a negative particle
current.
\end{enumerate}

As explained, fluctuations are essential for this mechanism, the source of
these usually being the temperature in most systems. However, the
temperature in our experimental system is too low to provide fluctuations of
sufficient amplitude. As previously stated, the role of temperature
fluctuations can also be played by a random distribution of pinning centers
which may add the needed stochasticity to vortex motion.

\begin{figure}[tbp]
\centering
\includegraphics[width=0.6\textwidth]{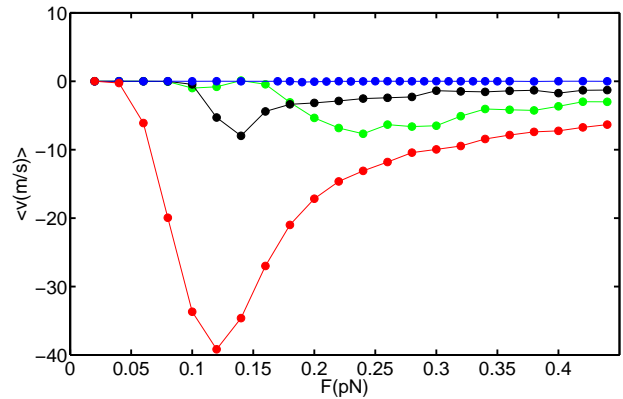}
\caption{Simulation results, average vortex speed versus amplitude of
applied force in the absence of intrinsic pinning, for $n=2$ (green), $n=3$ (black), $n=4$ (blue) and $n=6$ (red) vortices per unit cell. The line joining the points is only a guide for the eye. }
\label{fig_sindefectos}
\end{figure}

As expected, removing the pinning centers from simulations makes the
transverse signal disappear for $n=4$ and drastically reduces the signal for 
$n=2$ and $n=3$. For $n=4$ and without intrinsic pinning disorder the Abrikosov
lattice formed by the vortices approximately matches the triangle square
lattice, forming a well ordered almost triangular lattice, as explained and
depicted in \cite{dinisn4}. When pulled horizontally, vortices move in
almost perfect order from one column to the next, yielding a vanishing
vertical current, as shown in figure \ref{fig_sindefectos} (see also
supplementary material). For $n=2$, the vortex lattice is not perfectly
ordered even in the absence of intrinsic pinning disorder, presenting some
interstitial vortices, a small part of the triangles trapping just one
vortex or three. This disorder is enough to provide a small transversal
signal as shown also in figure \ref{fig_sindefectos}, however, the addition
of pinning disorder clearly enhances the signal, as also happens for $n=3$.

Finally, the effect or pinning disorder is the opposite for $n=6$, where the
vortex interaction is stronger and matching between the vortex and the array
of triangles is worse in the absence of intrinsic disorder. In this case,
adding intrinsic pinning merely increases the overall pinning and the
movement of the vortices is slower, giving less signal.

Supplementary multimedia material allows the reader to compare simulations
with or without intrinsic pinning disorder and inspect the mechanism behind
rectification.

\section{Conclusions}

In summary, hybrids of superconducting films with periodic asymmetric
nanotriangles show transverse ratchet effect, i.e.\ injecting an ac current
parallel to the reflection symmetry axis yields a dc output voltage in the
perpendicular direction. This effect can be modelled in the framework of
Langevin equation for interacting particles in 2 dimensions. The simulations
provide an intuitive mechanism for the observed transverse rectification of
vortices. Moreover, we have shown that intrinsic random pinning is necessary
to reproduce the experimental results. The role played by the intrinsic
pinning is smeared out increasing the number of vortices.

\ack

We acknowledge funding support by Spanish Ministerio de Ciencia e Innovaci%
\'{o}n grants NAN04-09087 and MOSAICO, FIS2005-07392, Consolider
CSD2007-00010, FIS2008-06249 (Grupo Consolidado) and CAM grant
S-0505/ESP/0337, and Fondo Social Europeo. Computer simulations of this work
were performed at the \textquotedblleft Cluster de c\'{a}lculo para T\'{e}%
cnicas F\'{\i}sicas\textquotedblright at UCM, funded in part by UE-FEDER
program and in part by UCM.

\end{document}